# Imaging Coherent Electron Flow

**B J LeRoy, A C Bleszynski, M A Topinka, R M Westervelt,
S E J Shaw and E J Heller**

Department of Physics, Division of Engineering and Applied Sciences and
Division of Chemistry and Chemical Biology

Harvard University, Cambridge, MA 02138 USA

**K D Maranowski and A C Gossard**

Materials Department

University of California Santa Barbara, Santa Barbara, CA 93106 USA

**Abstract.** Images of electron flow through a two-dimensional electron gas from a quantum point contact (QPC) can be obtained at liquid He temperatures using scanning probe microscopy (SPM). A negatively charged SPM tip depletes the electron gas immediately below it and decreases the conductance by backscattering electrons. Images of electron flow are obtained by recording the conductance as the tip is scanned over the sample. These images show angular patterns that are characteristic of electron flow through individual modes of the QPC, as well as well-defined branches at longer distances. The addition of a prism formed by a triangular gate bends electron paths as the electron density is reduced under the prism by an applied gate voltage. Under the conditions of the experiment, electron-electron scattering is the dominant inelastic process. By observing how the amplitude of backscattered electrons in images of electron flow decreases with added electron energy, we are able to determine the average length and time necessary for inelastic scattering. A dc voltage $V_o$ applied across the QPC accelerates electrons so that their energy is greater than the Fermi energy before inelastic scattering occurs. The signal is observed to decrease in amplitude and eventually disappear at distances from the QPC that decrease progressively as $V_o$ is increased.

## 1. Introduction

Scanning probe microscopy (SPM) is an important tool for understanding mesoscopic devices on a local scale. Recently it has become possible to image coherent electron flow spatially in a two-dimensional electron gas (2DEG) [1-4] opening the way for a more detailed understanding of 2DEGs and devices fabricated from them. This new understanding will be important for the design of devices that rely on the coherence of the electron, including implementations of quantum information processing. Previous experiments have measured the electron-electron scattering time in a 2DEG but obtaining spatial information has been difficult [5-7]. SPM techniques can provide more detailed spatial information about electron-electron scattering.

In this paper, we describe how scanning probe microscopy can be used to image coherent electron flow from a quantum point contact (QPC) in a 2DEG. Our previous work has shown that electrons flow in angular patterns that are characteristic of individual modes of the QPC at submicron distances [1] and form well defined branches at larger distances due to small angle scattering from charged donor and impurity atoms [2]. Interference fringes spaced by half the Fermi wavelength decorate the images demonstrating the coherence of electron flow. In this paper, we present images that show how electron paths are bent by a triangular gate that acts as an electrostatic prism. In addition, we use the imaging technique to probe the energy loss of electrons in a 2DEG due to electron-electron scattering. An additional dc voltage $V_o$ is applied across the QPC to accelerate electrons, increase the rate of electron-electron scattering, and decrease the distance they travel coherently. Measurements of the scattering time vs. $V_o$ agree with the theory of electron-electron scattering time in a 2DEG.

## 2. Measurement technique

Figure 1A illustrates the technique used to image electron flow. The samples are mounted in vacuum inside a scanning probe microscope, and both the sample and the SPM are cooled to a temperature of 1.7 K. A QPC is formed in a two-dimensional electron gas located 57 nm below the surface of an AlGaAs/GaAs heterostructure by two electrostatic gates on the surface. A negative voltage with respect to the 2DEG is put on the tip creating a small area below the tip that is depleted of electrons. This depleted area backscatters electrons passing through the QPC and reduces its conductance. When the tip is over an area of high electron flow there is a large reduction in the conductance, whereas when the tip is over an area of little flow there is a correspondingly small change in conductance. By measuring the QPC conductance G as a function of tip position, an image of electron flow is obtained. Figure 1B shows well defined conductance plateaus in steps of $2e^2/h$ of the QPC. The increase in conductance from one step to the next, as the width of the QPC is increased, represents the addition of a new quantum mode of electron flow through the QPC [8,9]. The inset in Fig. 1B is a scanning electron micrograph of the electrostatic prism device that consists of a QPC and a triangular gate located 1 μm to the left.

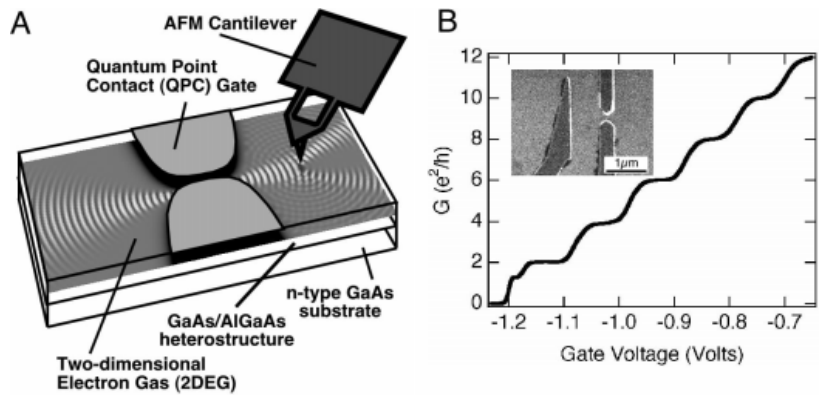

**Figure 1. A** Schematic diagram showing the measurement setup used to image coherent electron flow. **B** Quantum point contact conductance versus gate voltage showing well-defined conductance plateaus. The inset is a scanning electron micrograph of the electrostatic prism device.

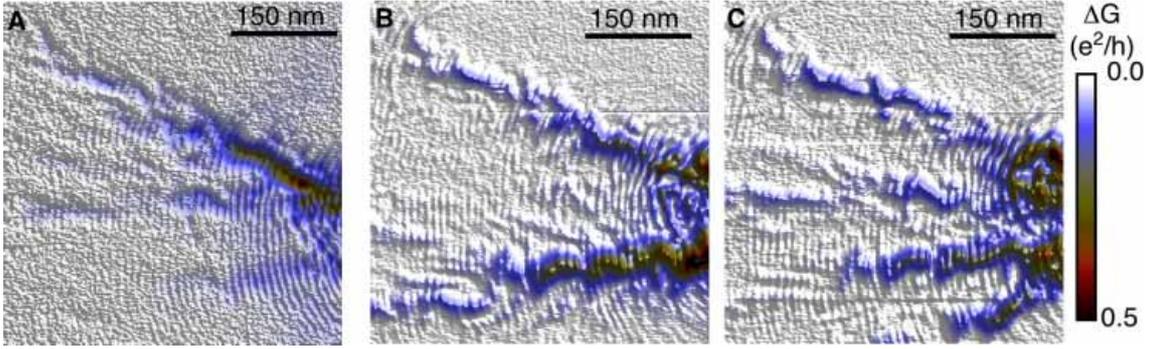

**Figure 2.** Images of coherent electron flow from the **A** first, **B** second and **C** third mode of a QPC located 700 nm past the right edge of the image. Dark areas have high electron flow and light areas have little or no flow.

Figure 2 shows a typical set of images of electron flow from the first three modes of a QPC. The light areas correspond to little or no conductance change and hence little electron flow. The dark areas have the largest change in conductance and the strongest electron flow. The center of the QPC is located $\cong$ 700 nm past the right edge of the images. In Figure 2A the conductance of the QPC is adjusted to the first conductance plateau and the flow comes from the QPC in one main lobe. As the QPC is opened a new lobe of current is added for each new mode. Figure 2B is an image of flow from the second mode obtained by subtracting the image of electron flow on the first plateau from the image on the second plateau. The electron flow on the second mode shows two main lobes of current. Figure 2C is an image of flow from the third mode, obtained in a similar manner, which has three main lobes. The number of lobes of current near the QPC is determined by the number of the mode [1].

All three images in Fig. 2 show branching as the electrons flow away from the QPC. This branching behaviour is due to the cumulative effect of many small angle scattering events [2]. These scattering events are caused by charged Si donor atoms located 22 nm from the 2DEG and by ionized impurities. In addition, the flow is decorated by interference fringes spaced by half the Fermi wavelength, that are oriented perpendicular to the direction of current. The spacing of the interference fringes provides a local measurement that can spatially profile the electron density in the 2DEG [3].

## 3. Electrostatic prism

The direction of electron flow can be bent by an electrostatic prism formed by a triangular gate on the surface of the sample. The voltage between the gate and the 2DEG changes the density of electrons underneath, changing their velocity. This is an electron analogue to Snell's law in optics [10,11]. An electron impinging on the edge of the gate at an angle $\theta_1$ to the normal will be bent to an angle $\theta_2$ after leaving. The relation between the two angles is given by

$$\frac{\operatorname{Sin}(\theta_1)}{\operatorname{Sin}(\theta_2)} = \left(\frac{n_2}{n_1}\right)^{1/2}$$

where $n_1$ is the electron density of the 2DEG, and $n_2$ is the density under the gate. As the voltage on the prism gate is made more negative, the density of electrons is decreased and the flow is bent upwards. This effect can be seen in Fig. 3, which consists of images of electron flow from the third plateau of a QPC for three different gate voltages. The dark lines on each image show the expected bending of electrons for the given image while the white lines show the expected bending for the other two voltages shown. The outline of the prism gate is shown by the grey triangle, which is located 1 μm to the left of the QPC (Fig. 1B). The electron paths bend upward as the voltage becomes more negative, in good agreement with theory. Interference fringes spaced by half the Fermi wavelength are located throughout the images showing that the flow is coherent. The prism can be used as a switch for coherent electrons since the direction of the electron beam can be changed with an applied gate voltage.

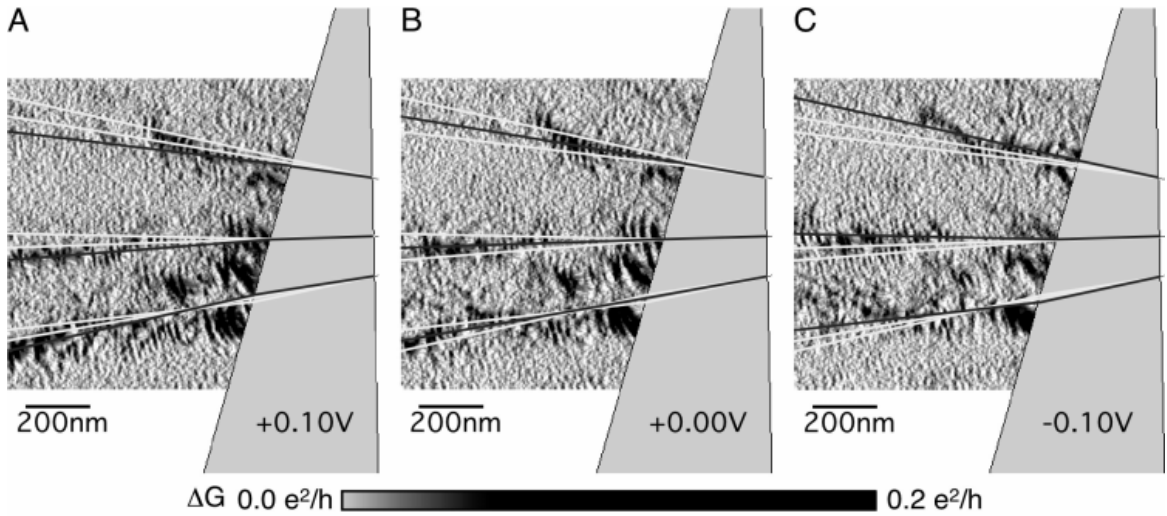

**Figure 3.** Three images of electron flow bent by the electrostatic prism, as the prism gate voltage is decreased from 0.1V to –0.1V, as indicated. The grey lines show the expected bending of electrons for the three prism gate voltages – the dark line is for the given image. The grey scale shows the change in QPC conductance induced by the charged tip.

**4. Imaging electron-electron scattering**

Images of electron flow can be used to probe the coherence of electrons. Figure 4 is a schematic plot of the bottom of the conduction band when a dc voltage $V_0$ is applied across the QPC. This accelerates electrons as they move through the QPC, increasing their kinetic energy. Measurements of the differential conductance $g = \partial I/\partial V$ of the QPC were made by adding a small ac voltage 0.2 mV to $V_0$. To produce an image, the SPM tip voltage must be sufficient to backscatter electrons with excess kinetic energy $V_0$. If the electrons lose energy in the round trip from the QPC to the tip, they will not be able to go back through the QPC and reduce its conductance. The images obtained in this way show the flow of coherent electrons that have not scattered with other particles.

Figure 5 shows the electron flow on the first conductance plateau for two different dc voltages $V_0$: for Fig. 5A, $V_0 = 0$ and for Fig. 5B, $V_0 = 2.4$ meV. The pattern of electron flow in both images is quite similar, indicating that the dc voltage does not change the paths that electrons take. A measure of electron scattering is given by the relative strength of flow in the images. Near the QPC the two images are quite similar. As the distance from the QPC increases, the intensity of coherent electron flow decreases more quickly with distance in Fig. 5B, indicating that electrons lose energy via scattering with other particles as they move away from the QPC.

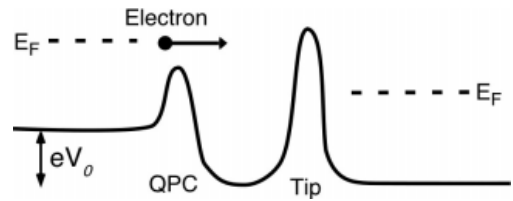

**Figure 4.** Schematic diagram that shows the effect of a dc voltage $V_0$ applied across the QPC. If electrons at $E_F$ coming from the left lose some of their energy before hitting the tip and returning to the QPC, they will not pass through the QPC and will not reduce the differential conductance.

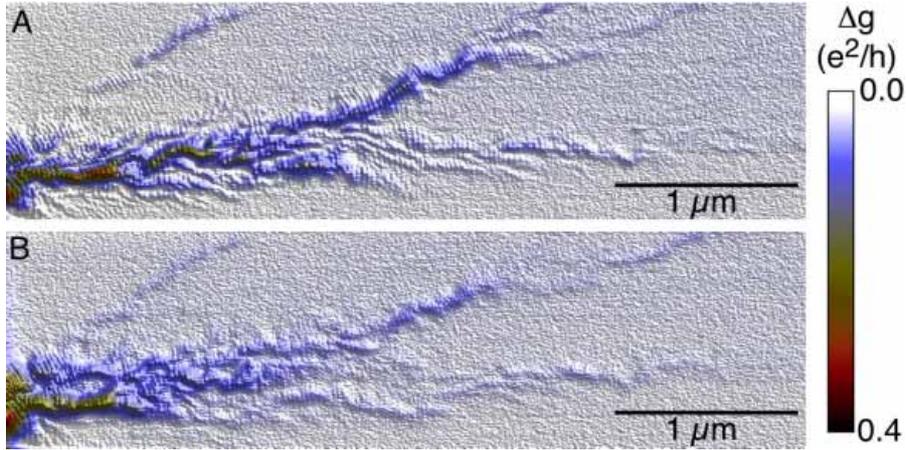

**Figure 5.** Two images of electron flow with applied voltages $V_0$ of **A** 0 meV and **B** 2.4 meV. The differential conductance decreases more quickly with distance in **B**, because electrons lose their excess kinetic energy through electron-electron scattering.

Figure 6 shows the electron flow for a small area of current flow located approximately 1 μm from the QPC for four applied voltages, (A) $V_0 = 0$ meV, (B) $V_0 = 1$ meV, (C) $V_0 = 2$ meV, and (D) $V_0 = 3$ meV. In each image, the width of the QPC was adjusted to pass only the first mode of conductance. These images demonstrate how the dc voltage $V_0$ reduces the coherence of electron flow. As $V_0$ increases, the electrons scatter more strongly with other particles and lose their excess kinetic energy more quickly, reducing the intensity of coherent electron flow in Fig. 6.

Figure 7 shows how the electron wavelength changes with the voltage $V_0$ applied across the QPC at two distances from the QPC: circles 1 μm and triangles 2 μm. As $V_0$ is made more positive, the electron wavelength increases at both distances, because the kinetic energy is decreased by the applied dc voltage. The measured wavelength is obtained from a Fourier transform of a one-dimensional slice of the data perpendicular to the fringes. The two solid lines in Fig. 7 plot the wavelength computed from the measured 2DEG density at $V_0 = 0$ and the dc voltage $V_0$. The measured wavelengths agree with theory, showing that the electron flow imaged by the differential conductance is coherent, and demonstrating that we are able to measure their energy.

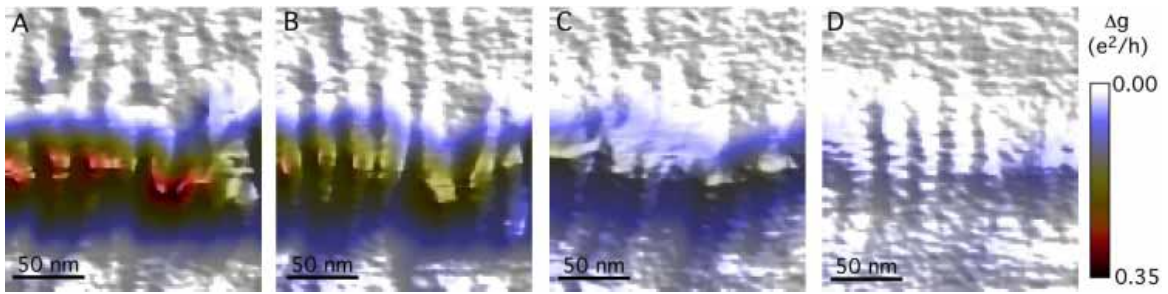

**Figure 6.** Four images of electron flow for increasing dc voltage $V_0$ applied across the QPC **A** 0 meV, **B** 1 meV, **C** 2 meV, and **D** 3 meV. The image disappears with increasing voltage, because the electrons scatter more frequently with larger applied voltage.

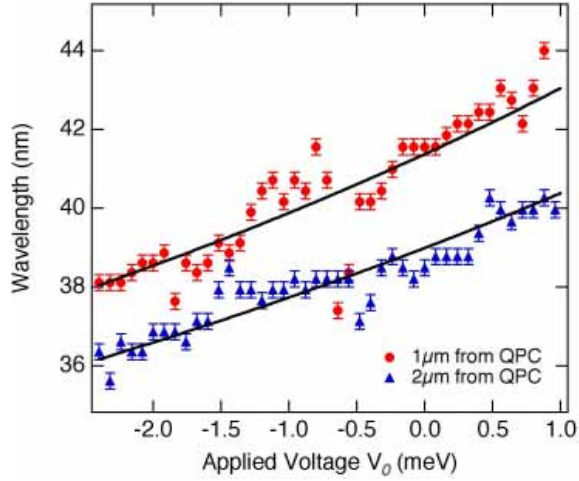

**Figure 7**. Measured electron wavelength as a function of applied dc voltage $V_0$ for two different distances from the QPC. The line is the expected wavelength based on the electron density and the additional energy provided by $V_0$.

Figure 8 is a plot of the change in differential conductance $\Delta g$ caused by the SPM tip vs. dc voltage $V_0$ applied across the QPC averaged over a 150 nm x 150 nm area. The curves have been normalized to 1 at $V_0 = 0$ meV, and they are offset vertically for clarity. As shown in Fig. 8, $\Delta g$ decreases as the kinetic energy is increased by $V_0$, because the scattering rate increases. Figure 8 also shows that $\Delta g$ decays more quickly with applied dc voltage at larger distances from the QPC. This occurs because electrons have more time to scatter during the roundtrip between the QPC and the tip, and they are more likely to lose some of their kinetic energy. At the distance 0.6 μm, $\Delta g$ decreases by 25% as the dc voltage $V_0$ increases from 0 to 2 meV. This decrease in $\Delta g$ with dc voltage becomes larger at larger distances: 45% at 1.0 μm and 50% at 1.4 μm.

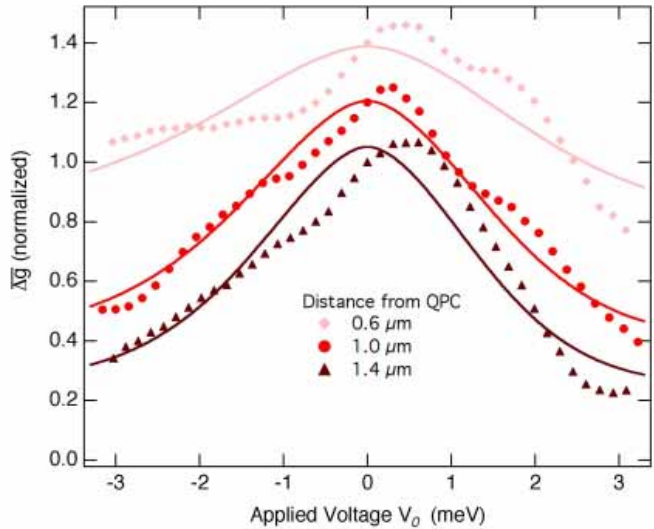

**Figure 8.** Plot of the average change in differential conductance $\Delta g$ caused by the SPM tip versus dc voltage $V_0$ applied across the QPC at the three distances from the QPC indicated in the figure. The solid lines are theoretical calculations based on the electron-electron scattering time. The curves are normalized to 1 at $V_0 = 0$ and are offset vertically by 0.2 for clarity.

At the temperatures and voltages used in this experiment, electron-electron scattering is the dominant energy loss mechanism. The theoretical work of Chaplik [12] and of Giuliani and Quinn [13] give the electron-electron collision rate at T = 0 K when a dc voltage $V_0$ accelerates electrons:

$$\frac{1}{\tau(V_0)} = \frac{E_F}{4\pi\hbar}\left(\frac{V_0}{E_F}\right)^2 \left(\ln\left(\frac{E_F}{V_0}\right) + \ln\left(\frac{2q_{TF}}{p_F}\right) + \frac{1}{2}\right)$$

Here $q_{TF} = 2me^2/\varepsilon\hbar^2$ is the Thomas-Fermi screening vector in two dimensions, $E_F$ is the Fermi Energy, $p_F$ is the Fermi momentum, m is the electron effective mass and $\varepsilon$ is the dielectric constant. If an electron undergoes a collision with another electron, then it does not contribute to the differential conductance image, as described above. The probability that an electron can travel a distance L without colliding with another electron is given by

$$\mathrm{Exp}\left(\frac{-L}{v\tau}\right)$$

where v is the velocity of the electron and $\tau$ is the time between collisions. The change in differential conductance $\Delta g$ should also vary as an exponential because we only measure electrons that have not undergone collisions. The solid lines in Figure 8 show the computed $\Delta g$ vs. $V_0$ using this model. The theory agrees with the data quite well, demonstrating that the imaging technique is able to measure energy relaxation, and that the mechanism is electron-electron scattering. The parameters used to calculate the curves are the electron density measured from the spacing of the interference fringes at $V_0 = 0$ and the distance the electron travels. The only free parameter is the height of the curve at $V_0 = 0$. The theory predicts a variation in $\Delta g$ that is symmetric about $V_0 = 0$, but the data are not perfectly symmetric. This is possibly caused by an asymmetry in elastic small angle scattering near the QPC caused by the gates and ionized donor and impurity atoms.

As a further comparison between theory and experiment, the 2DEG density can be varied using a back gate underneath the electron gas to change the electron-electron scattering rate. The points in Fig. 9 plot the measured change in differential conductance $\Delta g$ versus dc voltage $V_0$ applied across the QPC at a distance 1.0 μm for the three back gate voltages indicated. The solid lines plot the computed $\Delta g$ using the measured 2DEG density and the same roundtrip distance from the QPC. Applying a negative voltage to the back gate reduces the density of the 2DEG and the kinetic energy of the electrons. As shown in Fig. 9, the effects of electron-electron scattering at a given dc voltage $V_0$ are increased at lower 2DEG densities, making the signal $\Delta g$ decrease more quickly with $V_0$, in agreement with theory.

## 5. Conclusions

We have imaged coherent electron flow from a QPC in a 2DEG inside a GaAs/AlGaAs heterostructure at liquid He temperatures by using a scanning probe microscope with a charged tip. Near the QPC electrons flow in a well defined angular pattern - the number of lobes of current is equal to the number of the quantum mode of the QPC; one lobe for the first mode, two for the second, etc. At distances greater than ~ 1 μm from the QPC, the electron flow forms narrow branches caused by elastic small angle scattering. Interference fringes spaced by half the Fermi wavelength decorate all of the images, demonstrating the coherence of electron flow. Electron paths are bent by a triangular electrostatic gate that acts as a prism by reducing the electron density and velocity below; this triangular gate can be used as a switch for electrons. Inelastic electron-electron scattering in the 2DEG was imaged by applying a dc voltage across the QPC. Electrons accelerated by the dc voltage increase the electron-electron scattering rate, which decreases backscattering from the SPM tip. By measuring the strength of the backscattered signal as a function of the excess kinetic energy, the rate of electron-electron scattering is determined. The electron-electron scattering rate was measured at three distances between the QPC and the tip, and at three different 2DEG densities, changed using a back gate. In all cases, the electron-electron scattering rate is found to be in good agreement with theory. The ability to image electron-electron scattering is important for the design of devices that rely on the coherence of electrons.

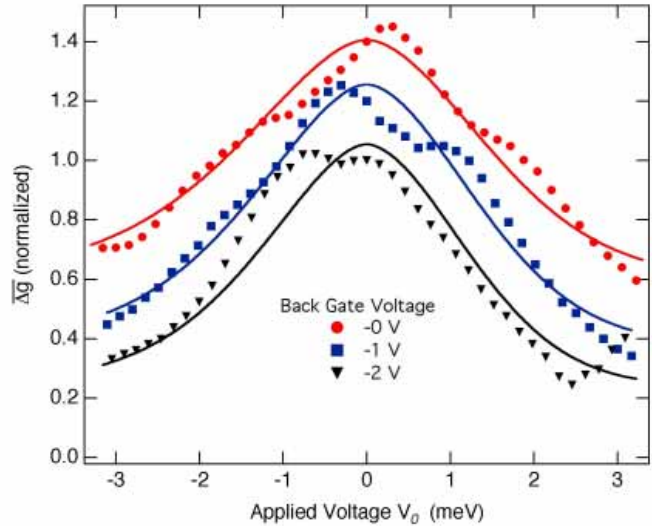

**Figure 9**. Plot of the average change in differential conductance Δg at a distance 1.0 μm between the QPC and the tip versus dc voltage $V_0$ applied across the QPC. The three sets of data are for different back gate voltages, as indicated, which change the 2DEG density. The solid lines are based on the computed electron-electron scattering rate. The curves are normalized to 1 at $V_0 = 0$ and offset by 0.2 from each other for clarity.